\documentclass{raa}            

\usepackage{graphicx,times}             
\usepackage{natbib}
\bibpunct{(}{)}{;}{a}{}{,}
\usepackage{amssymb,amsmath}
\usepackage[pagebackref=true]{hyperref}

\begin{document}

%
%

\title{The SVOM French Science Center Infrastructure}

\volnopage{Vol.0 (202x) No.0, 000--000}      
\setcounter{page}{1}          

\author{
    H. Louvin\inst{1}
    \and D. Corre\inst{1}
    \and A. Formica\inst{1}
    \and L. Jouvin\inst{1}
    \and K. Tazhenova \inst{1}  
    \and T. Sadibekova \inst{2} 
    \and J. Palmerio\inst{1}    
    \and N. Dagoneau\inst{1}    
    \and A. Sauvageon \inst{1} 
    \and A. Claret \inst{1}  
    \and F. Agneray\inst{3}
    \and C. Moreau\inst{3}
    \and T. Fenouillet\inst{3}  
    \and Y. Roehlly\inst{3} 
    \and J-C. Thome\inst{3} 
    \and L. Michel\inst{4}
    \and P. Maggi\inst{4}
    \and L. Kleiver\inst{4} 
    \and P. Maeght\inst{5}
    \and U. Jacob\inst{5}   
    \and F. Piron\inst{5}
    \and P. Bacon\inst{6}
    \and N. Bellemont\inst{6}
    \and C. Cavet\inst{6}   
    \and F. Daigne\inst{7}
    \and F. Lacreu\inst{7}
    \and G. Tcherniatinsky\inst{7}
    \and J. Wang\inst{7}
    \and Y. Canton\inst{8}
    \and I. Jegouzo\inst{9}
    \and N. Leroy\inst{10}
    \and S. Du\inst{10}
    \and S. Lion\inst{10}
    \and R. Le Montagner\inst{10}    
    \and C. Van Hove\inst{10}    
    \and J. Alaux\inst{11}
    \and M. Boiziot\inst{11}
    \and T. Auphan\inst{12}
}

\institute{
    CEA Paris-Saclay, Institut de Recherche sur les lois Fondamentales de l'Univers, 91191 Gif sur Yvette, France;\\
    \and Université Paris-Saclay, Université Paris Cité, CEA, CNRS, AIM, 91191 Gif-sur-Yvette, France;\\ 
    \and Aix Marseille Univ, CNRS, CNES, LAM, Marseille, France;\\ 
    \and Université de Strasbourg, CNRS, Observatoire astronomique de Strasbourg, UMR 7550, 67000 Strasbourg, France;\\ 
    \and Laboratoire Univers et Particules de Montpellier, Université Montpellier, CNRS/IN2P3, 34095 Montpellier, France;\\ 
    \and Université Paris Cité, CNRS, CEA, Astroparticule et Cosmologie, 75013 Paris, France; \\ 
    \and Sorbonne Université, CNRS, UMR 7095, Institut d'Astrophysique de Paris, 75014, Paris, France;\\ 
    \and LUX, Observatoire de Paris, Université PSL, Sorbonne Université, CNRS, 92190 Meudon, France;\\ 
    \and UNIDIA, Observatoire de Paris - PSL, CNRS, 92190 Meudon, France;\\
    \and Université Paris-Saclay, CNRS, IJCLab, 91405, Orsay, France;\\ 
    \and IRAP, Université de Toulouse, CNRS, CNES, Toulouse, France;\\ 
    \and Aix Marseille Univ, CNRS/IN2P3, CPPM, Marseille, France;\\ 
\vs\no
   {\small Received 2025 December 15; accepted 2026 March 17}
}

\abstract{
    At the heart of the SVOM French ground segment, the French Science Center is a cloud-based platform which provides services and tools for the management, storage, scientific processing and visualization of SVOM data for the French community. This digital center is a critical node of the SVOM system since it is the single point of access to SVOM scientific data for the French community and the only component of the ground system connected to the SVOM VHF network allowing near-time communication from the satellite. Scientific processing pipelines are fully integrated into its infrastructure, allowing the automated production of high-level scientific data and the generation and broadcast of alerts to the scientific community at large. The software components of the French Science Center are running 24/7 and thus require a high level of automation, which led to the development of dedicated software relying on modern technical solutions such as micro-services, application containerization, infrastructure-as-code and continuous integration and deployment. In this paper, we describe the FSC infrastructure design, technological choices, and the process of SVOM data ingestion, archiving, and automated processing by the FSC scientific pipelines. We present analysis on the FSC performances in terms of availability, amount of data processed as well as processing speed.
}

\authorrunning{H. Louvin et al.}            
\titlerunning{The SVOM French Science Center Infrastructure}  

\maketitle

%
%
\section{Introduction}
\label{sect:intro}
SVOM is a French-Chinese space-based mission dedicated to the detection, localization and multi-wavelength study of high energy transient astronomical events, with a focus on the most distant stellar explosions: gamma-ray bursts (GRBs). The mission is composed of two sets of instruments: the first is integrated on board a satellite, while the second is located on Earth \citep{cordier+2026+svom}.

\begin{figure}[h]
   \centering
   \includegraphics[width=\linewidth]{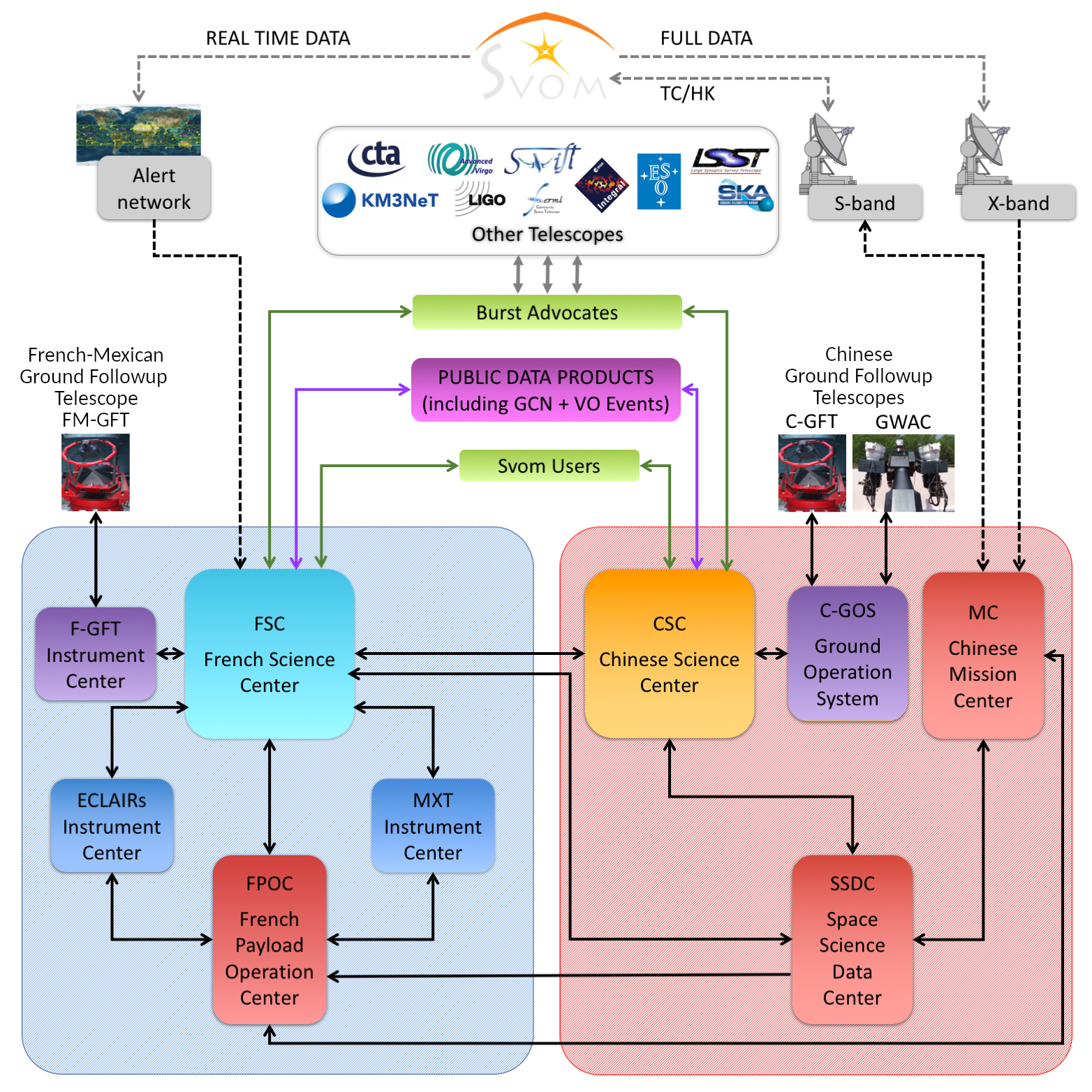}
   \caption{French-Chinese scientific ground segment}
   \label{fig:MGS}
\end{figure}

Figure~\ref{fig:MGS} shows the global organization of the SVOM mission ground segment elements. The French Mission Ground Segment, composed of the blocks shown in the left side of the figure, is divided in 6 main components:
\begin{itemize}
    \item the \textbf{Alert network}, the global network of VHF-band radio receivers allowing real-time reception of satellite data, specifically but not limited to gamma-ray bursts detection alerts \citep{cordier+2026+vhf};
    \item the \textbf{French Payload Operation Center} (FPOC), in charge of the management of the French payload;
    \item the three \textbf{Instruments Centers} (ICs), each dedicated to a French instrument, in charge of its monitoring, control, configuration and calibration:
    \begin{itemize}
        \item the ECLAIRs Instrument Center (EIC) for the ECLAIRs on-board telescope \citep{godet+2026+eclairs};
        \item the MXT Instrument Center (MIC) for the MXT on-board telescope \citep{gotz+2026+mxt};
        \item the French-Mexican Ground Follow-up Telescope (FM-GFT) Instrument Center (GIC) for the COLIBRÍ follow-up telescope \citep{basa+2026+fmgft}.
    \end{itemize}
    \item The \textbf{French Science Center} (FSC), which is the subject of this paper.
\end{itemize}

As the central node of the French Mission Ground Segment, the FSC is expected to carry out the following tasks:
\begin{itemize}
    \item Handle the reception and storage of the real-time monitoring and alert data coming from the VHF alert network in near-real time for all instruments \citep{cordier+2026+vhf};
    \item Handle the reception and storage of the complete observation and housekeeping telemetry data from the French instruments;
    \item Automatically produce high-level SVOM scientific data for all SVOM observation programs;
    \item Generate and broadcast alerts to the scientific community following the on-board detections of gamma-ray bursts (GRB) or other transient sources of interest;
    \item Provide services, interfaces and tools that allow the other SVOM centers, the members of the SVOM collaboration and the scientific community at large to visualize and access the SVOM data that they are privy to.
\end{itemize}

According to the french-side SVOM mission requirements, any software developed to perform these tasks must comply with technical requirements with regards to development workflow, versioning, continuous integration, code quality and security. The software infrastructure itself must provide high availability and storage replication. Performance-wise, the FSC is expected to distribute SVOM alerts coordinates to the SVOM ground-based telescopes in less than 30 seconds after detection for 65\% of the alerts and in less than 20 minutes for 95\% of the alerts.

The definition of the FSC technical perimeter started 9 years prior to SVOM launch. Since then, the development and maintenance of the FSC infrastructure and software components is under the responsibility of a team of approximately 40 software engineers and scientists distributed in eleven French national laboratories: APC, CEA/Irfu/DAP, CEA/Irfu/DEDIP, CPPM, IAP, IJCLab, IRAP, LAM, LUPM, LUX and ObAS.

The software solution that was chosen consists of a cloud-based platform centralizing all the software modules required to perform the FSC activities, including but not limited to: a database server, services handling raw data ingestion, the scientific processing pipelines producing high-added value scientific products, orchestration services handling the automation of data processing chains, and web-based visualization interfaces. This solution allows to reach the high level of automation required to meet the requirements and allows the SVOM collaborators to interact with ease with all the scientific data and tools available at FSC. The software infrastructure, the FSC components, their interactions and their performances are described in the following sections of this paper.

%
%
\section{FSC infrastructure}
\label{sect:infra}
\subsection{Architecture principles}

The French Science Center architecture is service-oriented: FSC applications are dedicated to single tasks, which might be of high complexity but are always as limited as possible in size and scope. All software components are fully containerized using Docker\footnote{\url{https://docs.docker.com/build/}} for easy deployment in the cloud, where they are orchestrated in clusters of Docker engines using Docker Swarm mode. There are three FSC cloud instances:
\begin{itemize}
    \item the \textit{integration} instance, where all code modifications are deployed continuously for testing;
    \item the \textit{pre-production} instance, where the new stable versions of software are tested in a controlled environment before being authorized to be put in production;
    \item the \textit{production} instance, where the latest FSC version is running and used by the scientific SVOM community.
\end{itemize}

\subsubsection{Infrastructure as code}
The FSC instances are deployed on OpenStack\footnote{\url{https://www.openstack.org/}} clouds provided by national research centers. The integration and pre-production instances are hosted by IJCLab (Irene Joliot Curie Laboratory in Orsay France) on their Virtual Data\footnote{\url{https://openstack.ijclab.in2p3.fr/activites/cloudvd/}} and the production instance on the cloud services of the French national research computing center CC-IN2P3\footnote{\url{https://doc.cc.in2p3.fr/en/Hosting/cloud-iaas.html}}.

The software infrastructures for all the FSC instances are provisioned and managed through scripts (rather than physical hardware configuration or interactive configuration tools). This process is known as "infrastructure as code" (IaC) and allows to fully automatize the creation of the software infrastructure for each of the FSC instances, including the creation, instantiation and configuration of the virtual machines (VMs), networks, security, storage, databases, etc.

This IaC approach is implemented at FSC using a combination of Terraform\footnote{\url{https://terraform.io}} and Ansible\footnote{\url{https://www.ansible.com/}} scripts. The former for provisioning the immutable OpenStack cloud infrastructure (Network, VMs, Volumes, Security Groups) and the latter for configuring the Virtual Machines.

\subsubsection{Software development}
The software modules composing the SVOM French Science Center were created and are maintained by development teams distributed in several French national laboratories: APC, CEA/Irfu/DAP, CEA/Irfu/DEDIP, CPPM, IAP, IJCLab, IRAP, LAM, LUPM, LUX and ObAS. In addition, the FSC integrates a data processing module that is provided by a team from NAOC, China. In order to ensure sharing, control and homogeneity among the productions of the different teams, the source code of all FSC software is saved and versioned on a GitLab\footnote{\url{https://about.gitlab.com/solutions/source-code-management/}} instance.

The number of programming languages used is limited: Python3\footnote{\url{https://www.python.org/}} for processing pipelines and most backend services, Java\footnote{\url{https://www.java.com/}} for raw data management modules, and JavaScript (React\footnote{\url{https://react.dev/}} and Angular\footnote{\url{https://angular.dev/}}) for web-based applications. In addition, C++ is also used as a built-in module in one of the Python library developed by the ECLAIRs trigger team, which allows to use methods extracted from the ECLAIRs onboard software.

We make an intensive use of the GitLab CI/CD\footnote{\url{https://about.gitlab.com/solutions/continuous-integration/}} continuous development tools, and follow a common git workflow as well as common continuous integration procedures to automatize the build, test, package delivery and deployment of FSC applications. All FSC software development teams are therefore using the same tools with the same configurations for:
\begin{itemize}
    \item code linting, i.e. proper formatting (pylint\footnote{\url{https://www.pylint.org/}}, ruff\footnote{\url{https://docs.astral.sh/ruff/}}, black\footnote{\url{https://pypi.org/project/black/}});
    \item unit testing (pytest\footnote{\url{https://docs.pytest.org/}}, jest\footnote{\url{https://jestjs.io/}});
    \item code quality and vulnerability scans (SonarQube\footnote{\url{https://www.sonarsource.com/products/sonarqube/}}, Trivy\footnote{\url{https://trivy.dev/latest/}});
    \item container building and pushing (Docker\footnote{\url{https://www.docker.com/}}, Kaniko\footnote{\url{https://github.com/GoogleContainerTools/kaniko}}, Harbor\footnote{\url{https://goharbor.io/}});
    \item documentation generation (Sphinx\footnote{\url{https://www.sphinx-doc.org/en/master/index.html}});
    \item changelogs generation (towncrier\footnote{\url{https://towncrier.readthedocs.io/}}).
\end{itemize}

\subsubsection{Data storage}
\label{sect:dataStorage}
Each FSC cloud instance has a dedicated PostgresSQL\footnote{\url{https://www.postgresql.org/}} database server, where all data received at FSC (i.e. raw telemetry and calibration data) and all data generated by FSC processings are stored. In addition to SVOM data, SQL databases are also used to store FSC software images, authentication data and software code quality profiles.
In addition to the SQL databases, a persistent storage disk space is used to save file-like data that are supposed to be kept during the whole lifespan of the mission and beyond. This space is stored on a Ceph file system\footnote{\url{https://ceph.io/en/}}.

The PostgreSQL and the persistent storage are backed up every 24 hours, and a replication of the PostgreSQL database is in place in another facility than the production site. The volume of data saved on FSC databases and persistent storage at the time of this paper's redaction on September 16, 2025 is presented in Table~\ref{table:Storage}.

\begin{table}
    \begin{center}
    \caption[]{FSC Data Usage (estimated 2025-09-16). File-like objects are stored on disk, while all other data is stored in SQL databases.}\label{table:Storage}
        \begin{tabular}{lrr}
            \hline\noalign{\smallskip}
            Data type                                           & DB size & Disk usage\\
            \hline\noalign{\smallskip}
            Science data (SDB)                                  & 905 MB  & 6.1 TB \\
            X-Band raw data                                     & 609 GB  & 1.2 TB \\
            VHF raw data                                        & 102 GB  & 31 GB  \\
            Instruments configurations and \\ housekeeping data & 94 GB   & 83 MB  \\
            Instruments calibrations                            & --      & 99 MB  \\
            SVOM detections alerts notices                      & 30 MB   & --     \\
            \hline\noalign{\smallskip}
            System monitoring data                              & 87 MB   & 190 GB \\
            Scientific pipelines orchestrator                   & 4.6 GB  & --     \\
            Burst Advocate web interface                        & 2.6 GB  & 1.1 GB \\
            \hline\noalign{\smallskip}
            Applications docker images                          & 926 MB  & 130 GB \\
            FSC Authentication                                  & 6.5 GB  & --     \\
            Code quality analysis data                          & 327 MB  & --     \\
            \noalign{\smallskip}\hline
        \end{tabular}
    \end{center}
\end{table}

\subsection{Centralized messaging}
\label{sect:messaging}
The FSC is composed of a multitude of distinct services. Those services have two ways of communicating with each other, depending on the number of services involved in the communication:
\begin{itemize}
    \item one-to-one communications are performed through REST API interfaces using standard HTTP protocol;
    \item one-to-many communications are made using a centralized messaging system.
\end{itemize}
For one-to-many communications, the FSC centralized messaging system is a publish/subscribe message broker hosted at FSC and relying on the message-oriented middleware NATS\footnote{\url{https://nats.io/}}.

NATS is an open-source publish-subscribe messages broker:
\begin{itemize}
    \item publishing services send messages to the server on pre-defined subjects identified by a string;
    \item subscribing services are connected as listeners to one or several subjects;
    \item the server broadcasts all messages received on a subject to all subscribers of that subject.
\end{itemize}

In order to ensure the distribution of messages to subscribers even if they are not connected at the time the message was sent, the FSC NATS server makes use of the built-in distributed persistence layer of NATS called JetStream\footnote{\url{https://docs.nats.io/nats-concepts/jetstream}}. This configuration allows the FSC services subscribed to the NATS messaging system to be safely restarted, for example when deploying a software update. All notifications missed during the service downtime will be recovered upon restart.

The Chinese Science Center uses the same architecture, but with a different message-oriented middleware called RabbitMQ\footnote{\url{https://www.rabbitmq.com/}} that relies on the MQTT protocol \citep{huang+2026+csc}. The messaging systems are key components of the interface between FSC and the Chinese centers. A couple of services running at FSC are connected to the CSC MQTT server in order to be notified about the availability of new data (more on that in \ref{sect:xband}).

%
%
\section{Raw data management}
\label{sect:rawData}
\subsection{VHF raw data management}
\label{sect:vhf}
The FSC is the only SVOM component connected to the VHF alert network, which allows it to receive and analyze limited-size communications from the satellite in near real-time \citep{cordier+2026+vhf}. The other SVOM centers such as the Instrument Centers or Chinese centers do not have a direct connection to the VHF network and therefore retrieve VHF data from the FSC.

The VHF data stream consist of 94-bytes encoded messages or "packets", broadcasted every 1.83 seconds from the on-board VHF radio transmitter. The messages are received on the ground by a network of VHF radio antennas which demodulate the signal and send the digitized binary messages content to the FSC through a REST API. The packets are received by the so-called "vhf-manager" service, which saves them in a dedicated database. The VHF raw data ingestion entails several steps:
\begin{enumerate}
    \item filtering of duplicated messages, which are due either to the simultaneous reception by multiple antennas of a single packet or to the satellite repeating the broadcast of the same packet several times;
    \item validation of the packets integrity via checksum verifications;
    \item unpacking of the binary VHF packet content using a dedicated FSC service that is responsible for reading the binary VHF packets and converting them in JSON format;
    \item computing of on-ground unified identifiers depending on the packets content (GRB-related data only);
    \item saving of the interpreted VHF packets contents in the VHF database;
    \item broadcast of notifications regarding the availability of new VHF data to other FSC components using the centralized messaging system (see \ref{sect:messaging}).
\end{enumerate}

In order to improve the reliability of SVOM communications for alert-related data, some of the messages sent through the VHF network are also downloaded using the Chinese navigation satellite system BeiDou. This communication subsystem has a slower rate than VHF and a much smaller bandwidth but it allows to speed up the on-ground reception of alert data in some cases, and could even be vitally important in case of VHF network temporary unavailability.

The FSC provides an authenticated REST API which allows French and Chinese SVOM centers as well as authorized SVOM collaborators to access the VHF data stored in the VHF database.

\subsection{X-Band raw data management}
\label{sect:xband}
SVOM uses X-Band communications to download the complete observation and housekeeping telemetry data from the satellite. Unlike SVOM VHF telemetry, the X-Band telemetry reception and distribution is under the responsibility of the Chinese National Space Science Data Center (SSDC). The SSDC receives all telemetry data that the satellite dumps through X-Band antennas several times per day - every 2 to 6 hours depending on the satellite orbit. The telemetry concerning French instruments is then uploaded to a dedicated FTP server.

The availability of French payload X-Band data is notified to FSC through the MQTT messaging server (see \ref{sect:messaging}). Upon reception of the notification by the dedicated service at FSC, the X-Band telemetry data is downloaded from the SSDC server and sent to the "xband-manager" service for storage and processing. The data files are stored on the persistent data storage file system, and their metadata are saved in a dedicated database.

The X-Band files that are received at FSC contain chains of 1022-bytes binary messages, which are extracted from each file by the xband-manager. Due to the sheer volume of data that is received through the X-Band channel, it is inconceivable for the xband-manager to interpret the data on the fly and store JSON values in the database in the same way that it is done for VHF data. Only the first few bytes of each X-Band packet are decoded to extract metadata (such as the type of data, the origin instrument, the time of packet production and the current observation number) after which both the metadata and the binary packet are saved in the database, where all the other FSC components can safely access them through a authenticated REST API. Notifications of availability are sent through the centralized messaging system (see \ref{sect:messaging}) so that the automated X-Band data processing can start.

\subsection{Ancillary data management}
\label{sect:crest}
In addition to the raw SVOM instrument data, the SVOM ground centers also exchange mission ancillary data files (mostly in JSON, CSV and TXT formats). These files include the mission observation work plan, statistics on executed observations, predicted ephemerids for the spacecraft as well as derived information such as planned X-Band passes and South Atlantic Anomaly flybys.

Such files are regularly sent to FSC by the French Payload Operations Center (FPOC) using CREST, an FSC-hosted service in charge of their reception, archiving and provision. CREST is a file-indexing database allowing for an accurate tracking of time-dependent data, with associated REST API for uploading and downloading files. The codebase of this service originates from a development in the High Energy Physics community for the management of Conditions Data \citep{laycock+2018}.

The FSC CREST service is also used as the main file exchange interface between the French Instrument Centers and FPOC for instrument configuration files, health and alarm reports, on-board software updates, and more.

%
%
\section{Scientific data processing}
\label{sect:dataProcessing}
Scientific data processing at FSC is performed by dedicated services referred to as "pipelines". There are currently 13 data processing pipelines running continuously at FSC, totalizing 73 execution modes and about 180 different output types. In this section we describe how these services are interfaced with the other parts of the system and how they are automatically triggered to process data in real-time, with a focus on the alert processing chain and its performances. The inner workflow and data analysis algorithms of the highest-level scientific pipelines operated at FSC are described in dedicated papers \citep{goldwurm+2026+ecpi,piron+2026+eclgrm,maggi+2026+mxtpipes,wu+palmerio+2026+vvppvtac}.

\subsection{Data models}
\label{dataModel}
The FSC workflow comprises many modules that have been developed separately by different contributors. These modules must be able to communicate with each other and process the scientific products that circulate within the infrastructure. To ensure consistency in such a heterogeneous architecture, a high level of interoperability is achieved between components at three levels, all of which are based on JSON resources \citep{michel+2020+jsonschemes}.
\begin{itemize}
     \item \textbf{vocabulary} : All properties contained in the messages exchanged by different modules are named according to a common vocabulary defined in a JSON schema.
     \item \textbf{message format} : Messages exchanged via NATS/JetStream or REST endpoints are validated against specific JSON schemas.
     \item \textbf{product descriptors} : All FITS files that may be stored in the science database are described in JSON files. These descriptors provide accurate descriptions of quantities that can be used to retrieve specific products as well as various metadata. Each individual FITS file is validated against its descriptor before being ingested by the science database and any discrepancy interrupts the ingestion process.
\end{itemize}
In addition, most services expose a common REST API, as detailed in \ref{bricks}.

\subsection{Science products database}
The Science Database (or SDB) is the database storing all scientific products - in FITS format \citep{pence+2010+fits} - and related ancillary files produced by the FSC scientific pipelines for all SVOM observation programs. The SDB provides REST APIs allowing to import and export scientific products.

New data is sent to the SDB by the FSC pipelines using the SDB import API, which includes REST access points for importing new products and for updating existing ones. All submitted products are analyzed before being recorded, in order to validate their format and version according to the corresponding product data model (see \ref{dataModel}). Ancillary files can only be imported if they are properly associated to data products already present in the database. Whenever a scientific product is successfully imported into the SDB, its availability is notified to the other FSC services using the NATS/JetStream centralized messaging system.

The data provision of the SDB relies heavily on the Astronomical Information System (ANIS) \citep{agneray+2019+anis}, a generic tool aimed at facilitating and homogenizing the implementation of astronomical data of various kinds and catalogs in dedicated Information Systems. ANIS includes a REST API allowing to the FSC scientific pipelines to extract data from the SDB according to various search criteria, and a dedicated web interface through which SVOM collaborators can search, visualize and download data from the SDB.

\subsection{Calibration database}
The FSC Calibration Database (CalDB) is built upon a regular HEASARC dalibration database instance for FITS files \citep{george+2005+heasarc} extended with a specific capability for supporting YAML files in certain conditions. It provides a REST API to import and export SVOM calibration files, and a dedicated web interface allowing SVOM collaborators to query it.

The HEASARC’s calibration database system stores and indexes datasets associated with the calibration of high energy astronomical instrumentation. The system can be accessed by users and software alike to determine which calibration datasets are available, and which should be used for data reduction and analysis. The FSC CalDB instance is restricted to a single mission - SVOM - and 4 instruments - ECLAIRs, MXT, GRM and VT.

In addition to the regular FITS file storage, the SVOM instance of the CaldDB can also manage YAML files. YAML are all stored in a specific node named onboard (one per instrument). The use case for the YAML storage is to trace the configuration files that have been uploaded onboard and used by the onboard software. It is not possible for Instrument Centers (ICs) to predict the moment at which a given configuration is actually applied on board. The ICs have to rely on the telemetry to know when the on-board configuration changes. When it is the case, the IC operators upload the corresponding YAML to the CalDB with a time stamp corresponding to the on-board application time.

\subsection{Data processing automation}
Scientific data processing at FSC is for the most part automated. The sequential execution of the various processing pipelines composing the chains is under the supervision of the so-called orchestrator service. All FSC services communicate with each other either via direct REST API calls or through notifications sent using the centralized messaging server described in \ref{sect:messaging}.

\begin{figure*}[h]
   \centering
   \includegraphics[width=0.8\textwidth]{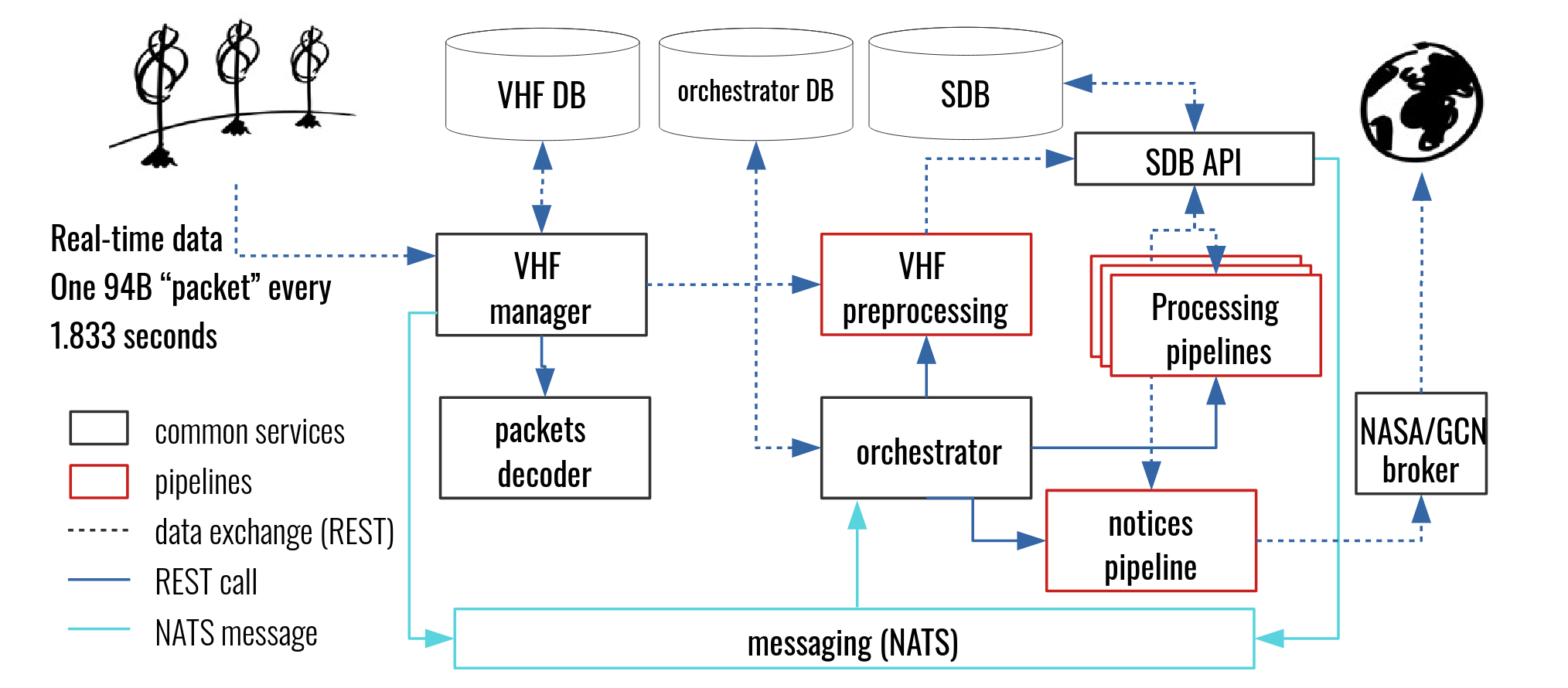}
   \caption{Complete Automated FSC Processing Chain for SVOM VHF Alert Telemetry, from the VHF Telemetry Reception to the Final Scientific Products and Alert Broadcast}
   \label{fig:vhfDataFlow}
\end{figure*}

\begin{figure*}[h]
   \centering
   \includegraphics[width=0.8\textwidth]{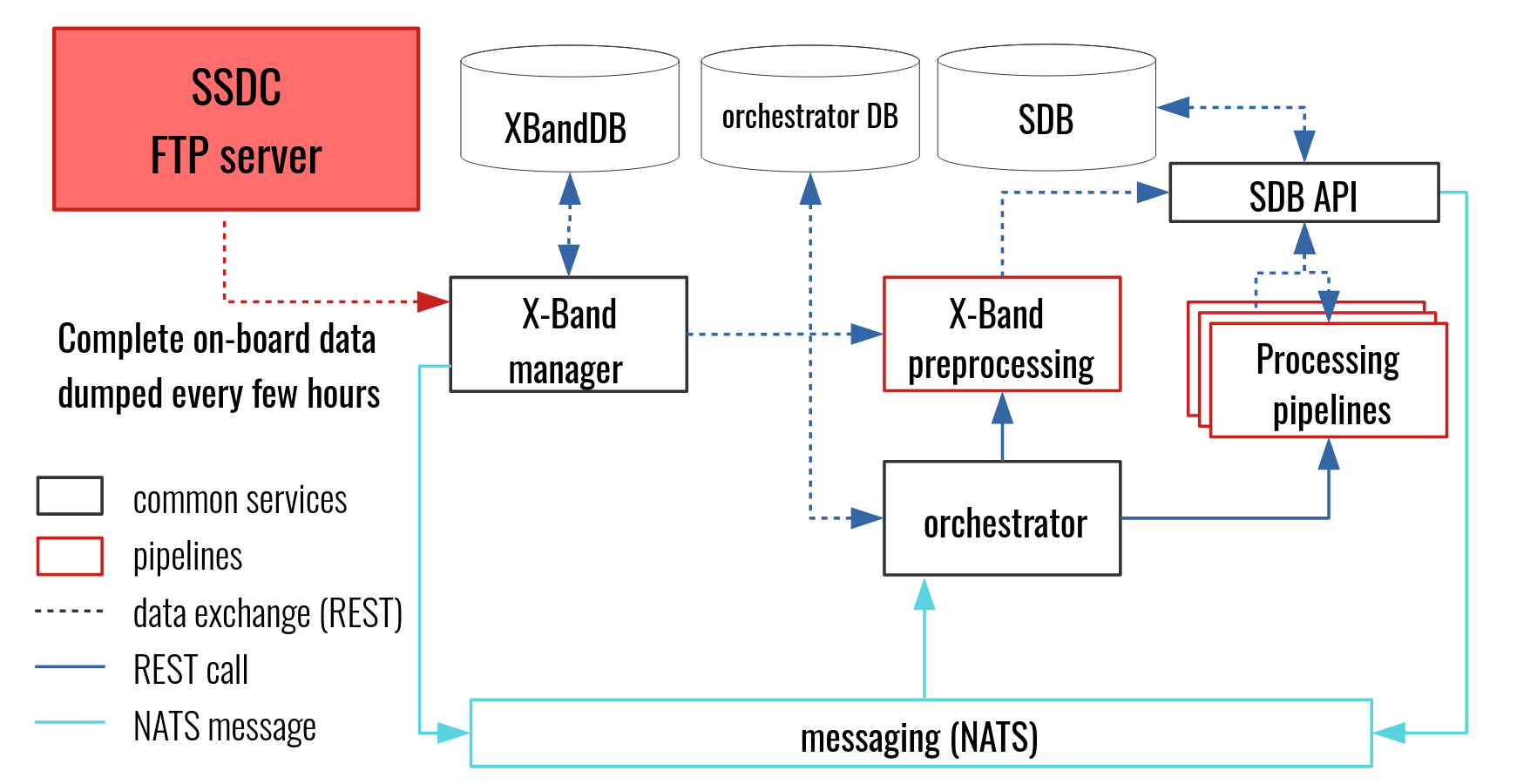}
   \caption{Complete Automated FSC Processing Chain for SVOM X-Band Telemetry, from the X-Band Telemetry Reception to the Final Scientific Products}
   \label{fig:xbandDataFlow}
\end{figure*}

The complete sequence of services involved in the automated processing of VHF telemetry data is schematically shown in Figure~\ref{fig:vhfDataFlow}. Figure~\ref{fig:xbandDataFlow} displays the same diagram for the automated processing of X-Band telemetry data.

\subsubsection{Data processing pipelines}
\label{bricks}
The FSC data processing pipelines are the FSC services responsible for processing SVOM scientific data and generating output products. These output products are most of the time FITS files that are sent to the Science Database (SDB) for storage when they are produced. All the FSC pipelines are continuously running services in the FSC cloud. They are mostly idle and will start processing data upon reception of a processing request, either from the automated FSC orchestrator or by a FSC user that can request a processing manually through the FSC website.

In order to ensure that all pipelines and their REST API are consistent with the common data model (see \ref{dataModel}), a generic task scheduler has been developed at FSC under the name "pipeline-bricks". All FSC pipelines use this scheduler and the associated REST API as a wrapper around the actual pipeline software. The pipeline-bricks principles are as follows:
\begin{itemize}
    \item One specific bricks instance always runs the same task sequence (namely the workflow).
    \item This sequence can be marginally modified in reprocessing mode.
    \item Each task must be made available as an individual executable command. The underlying programming language has no impact.
    \item Bricks instances can only run one sequence at the time. Any processing request received while a sequence is running will return an error.
    \item Tasks are executed one by one without parallelization.
    \item A processing sequence executes all tasks of the workflow. There is no conditional execution.
    \item Bricks is stateless in the sense that it has no memory of previous processing sequences or knowledge of the global context.
\end{itemize}

Bricks has no control on the task activities or the files that are imported or exported. It can however get some information from a runtime file that is shared between all tasks, provided that the core software writes information into it. This mechanism can allow bricks for example to provide through its REST API the list of generated science products for a given processing. This also allows bricks to automatically add VO data in some science products.

\subsubsection{Orchestrator service}
In order for data to be dynamically processed when available, automated processing requests must be sent to FSC pipelines through the pipeline-bricks REST API. The choice has been made to centralize these executions in a dedicated service developed at FSC, dubbed the FSC orchestrator. This central service is in charge of:
\begin{itemize}
    \item reacting when new data is available to be processed;
    \item sending processing request(s) to the pipeline(s) in charge of processing the data;
    \item handling processing queues if a pipeline is currently busy processing other data;
    \item keeping track of which pipeline has been triggered, when and with which data;
    \item monitoring ongoing processings, retrieving their logs and storing them in a database;
    \item updating processings logs and final status when processings are over.
\end{itemize}

Whenever new data is available in any of the various databases at FSC, a data notification is broadcasted on a specific NATS/JetStream subject with all the information required for the orchestrator to identify the type of data and the scientific observation to which it relates. Each database in the FSC ecosystem has its own publication subject on the NATS/JetStream server. A dedicated orchestrator service is connected to each of those message streams. Upon reception of a data notification, the corresponding orchestrator:
\begin{enumerate}
    \item determines if one (or several) processing operation(s) should be launched to treat the new data;
    \item sends the corresponding processing request(s) to the relevant pipeline(s) or service(s);
    \item monitors the evolution of the launched processing(s);
    \item takes actions upon the processing(s) termination: return status and logs retrieval, retry scheduling, etc.;
    \item stores relevant data concerning the processing(s) in a database.
\end{enumerate}

Most orchestrated processings terminate by sending processed data to the SDB or another FSC database. If and when this happens, the corresponding database sends a data notification on the NATS/JetStream server which might in turn trigger a new orchestrated processing. This allows to sequentially execute processings with a fine-grain monitoring on their inputs and outputs, producing higher level scientific products each step of the way.

\begin{figure}[h]
   \centering
   \includegraphics[width=\linewidth]{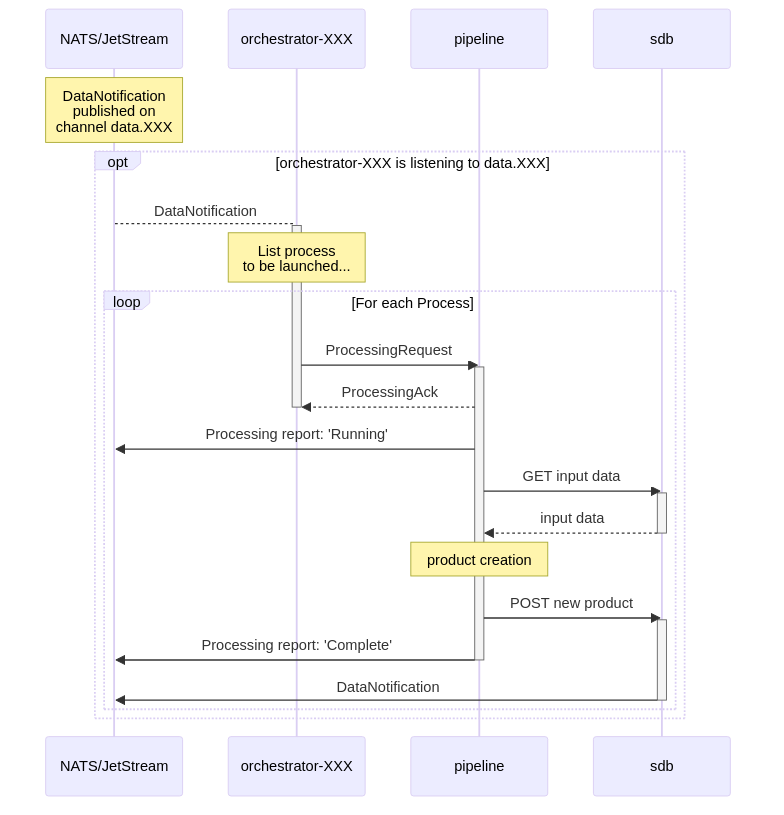}
   \caption{Orchestrated Pipeline Processings Flowchart}
   \label{fig:orchestratorDiagram}
\end{figure}

The pipeline execution process by the orchestrator services is illustrated in Figure~\ref{fig:orchestratorDiagram}.

\subsection{Alert data processing}
\label{sect:alertProcessing}
The so-called "core" observation program (CP) of SVOM is dedicated to Gamma-Ray Bursts (GRB) detection and study. GRBs are astronomical transients that can be very short lived, and the global scientific community must be warned as soon as possible of the detection of such phenomena in order for their study to be as extensive as possible. The on-board detection of GRBs by either one of the SVOM instruments having triggering abilities is received on the ground through so-called "alert" VHF messages providing location and other information on the detection. After that, "alert sequence" VHF messages are sent by one or more of the SVOM instruments containing more detailed observation data of the event.

Since the FSC is the only SVOM center directly connected to the VHF network, the processing pipelines that are dedicated to real-time analysis of SVOM VHF alerts are hosted at FSC. One of the responsibilities of FSC is to automatically generate "notices" in VOEvent format regarding the SVOM GRB detections and broadcast them to the worldwide scientific community in near-real time. The alert data processing at FSC consists of the following steps:
\begin{enumerate}
    \item the VHF manager receives the alert message through the VHF network;
    \item the packets decoder service decodes the alert message content;
    \item the VHF manager sends a notification on NATS/JetStream to signal availability of the alert data;
    \item the orchestrator receives the notification and triggers a first set of pre-processing pipelines;
    \item the pre-processing pipelines generate the first scientific products and import them in SDB;
    \item the SDB saves the products and sends a notification on NATS/JetStream to signal their availability;
    \item the orchestrator receives the notification and triggers the notices generation pipeline;
    \item the notices pipeline generates the alert notice and send them to the alert manager service;
    \item the alert manager service saves the notice in a dedicated database;
    \item the alert manager broadcasts the notice to the follow-up telescopes and the worldwide community using the NASA GCN network.
\end{enumerate}

Overall, a dozen of distinct FSC services are involved in the processing of alert data and the generation and broadcast of the VHF alert notices. The public SVOM VOEvent notices are broadcasted using the GCN network. Details can be found on the GCN website\footnote{\url{https://gcn.nasa.gov/missions/svom}} and all public SVOM VOEvents are available on the FSC public alerts page\footnote{\url{https://fsc.svom.org/alerts/}}.

\subsection{Statistics and performances}

\subsubsection{Data processing statistics}

A few statistics on the quantity of data received and produced at FSC are gathered in Table~\ref{table:FSCStats}. The numbers presented in this table cover the period from the launch on June 22, 2024 to September 30, 2025 with the exception of VOEvent notices production which only started in January 2025. The corresponding volume of data in bytes can be found in Table~\ref{table:Storage} discussed in Section~\ref{sect:dataStorage}.

\begin{table*}
    \begin{center}
    \caption[]{FSC data processing statistics from 2024-06-22 (SVOM launch) to 2025-09-30}\label{table:FSCStats}
        \begin{tabular}{lrr}
            \hline\noalign{\smallskip}
            Data type                                         & Count since launch & 2025 monthly average \\
            \hline\noalign{\smallskip}
            VHF Packets received and processed                & 30,645,621 & 1,862,060   \\
            X-Band raw data files received                    & 31,850 & 2,242 \\
            X-Band raw packets processed                      & 883,039,648 & 64,927,121 \\
            Instruments housekeeping telemetry data           & 158,666 & 10,541  \\
            \hline\noalign{\smallskip}
            Orchestrated processings launched                 & 405,761 & 31,115  \\
            Scientific products generated                     & 938,250 &  74,019 \\
            VOEvent notices generated (since 2025-01)         & 2,420 & 242 \\
            VOEvent notices public broadcasts (since 2025-01) & 447 & 45 \\
            \noalign{\smallskip}\hline
        \end{tabular}
    \end{center}
\end{table*}

The second column of Table~\ref{table:FSCStats} shows the number of items received or generated between June 22, 2024 and September 30, 2025. The rightmost column lists the monthly average since the end of the FSC commissioning period on January 1st, 2025. Since then, the system has been running continuously and processing data at a high rate, with on average 31 thousand orchestrated processings generating 74 thousand scientific products every month. The efficiency of the data processing performance is studied in the following section.

\subsubsection{Data processing performance}
As presented in section \ref{sect:alertProcessing}, the alert processing at FSC involves a lot of services among which the centralized messaging system, the orchestrator services and several pipelines and databases. Therefore, the study of the FSC alert processing performances allows to evaluate the overall FSC architecture efficiency.

\begin{figure}[h]
   \centering
   \includegraphics[width=\linewidth]{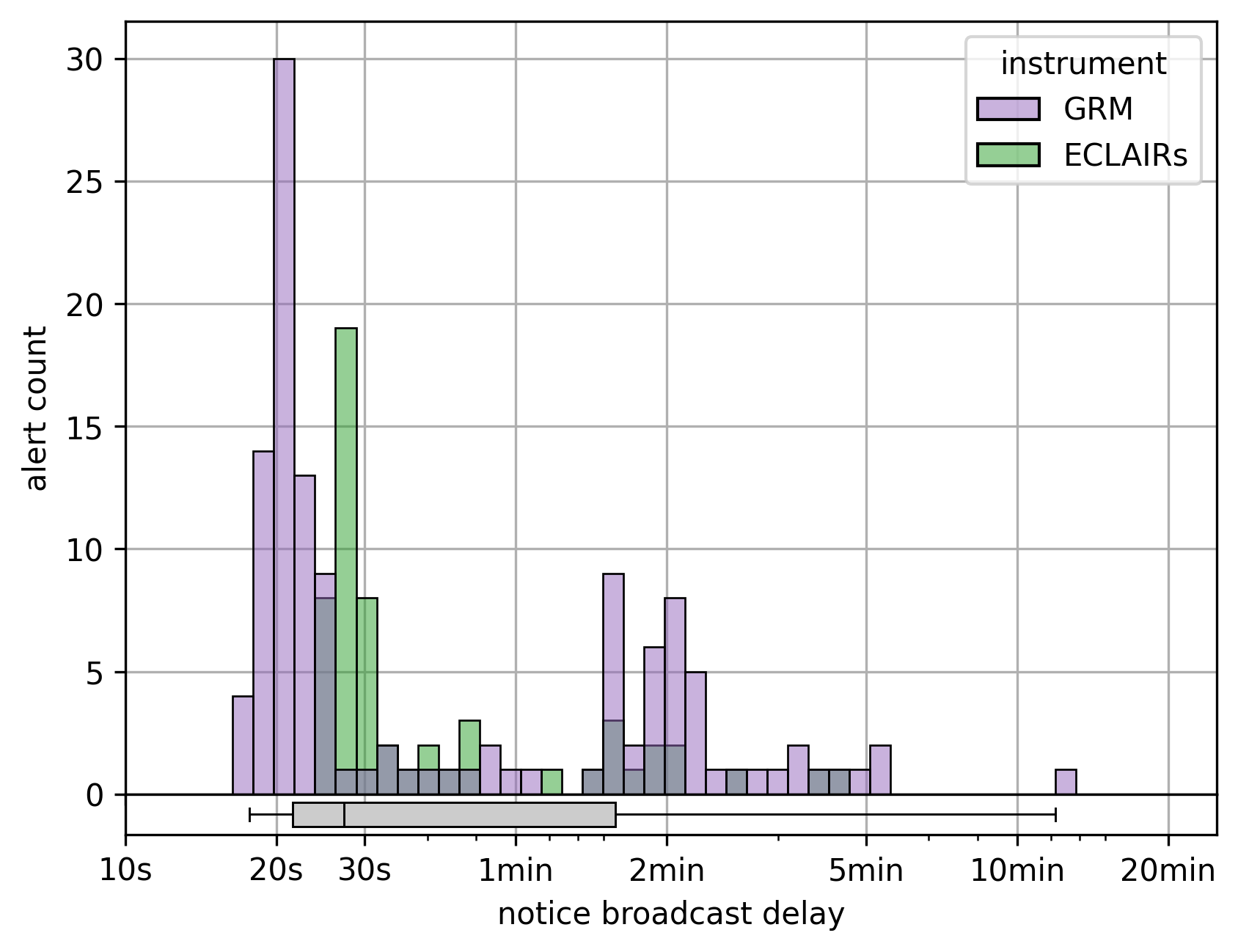}
   \caption{SVOM notices broadcast delays for the alert of Sept. 2025}
   \label{fig:notices_broadcast_delay_by_instrument}
\end{figure}

\begin{figure}[h]
   \centering
   \includegraphics[width=\linewidth]{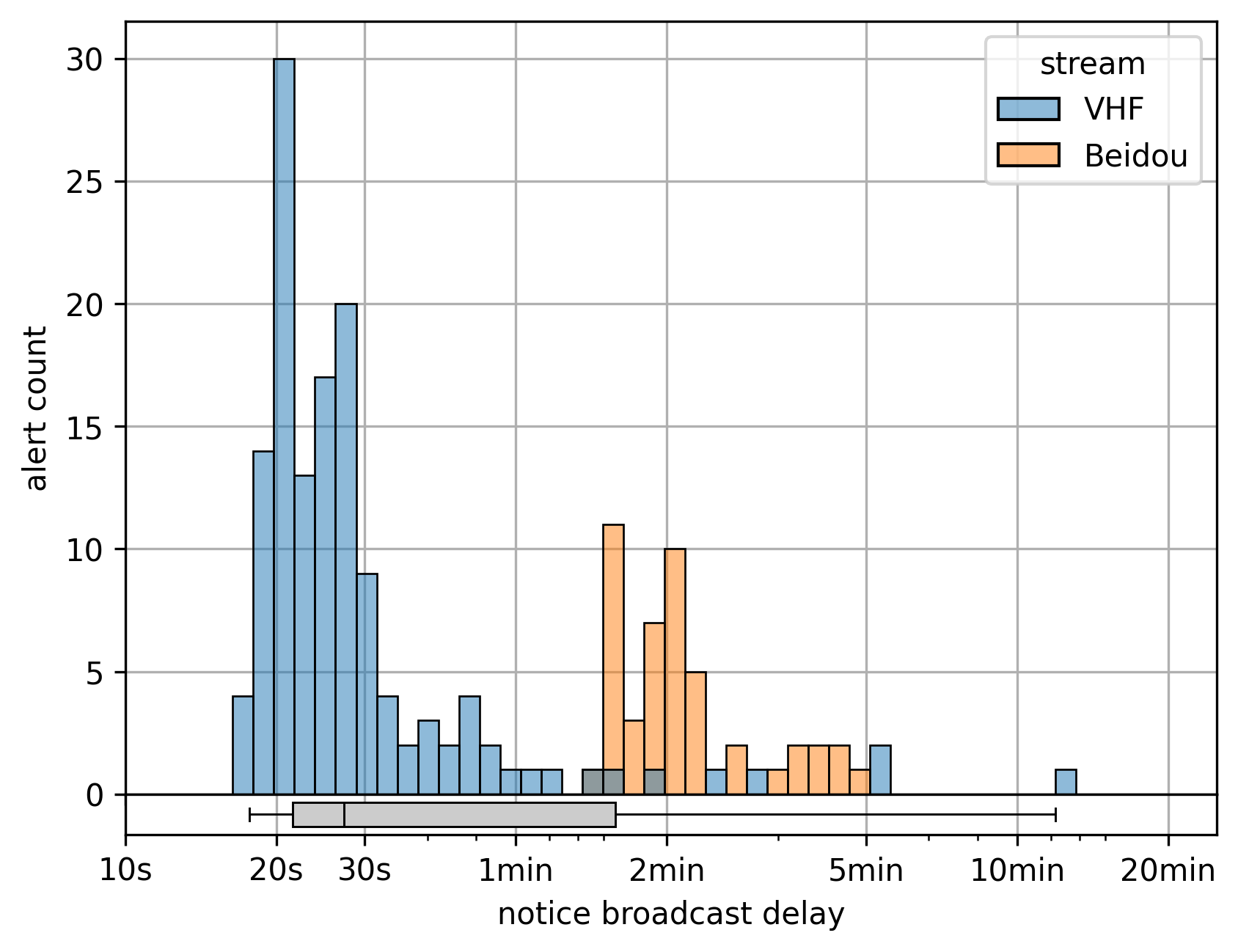}
   \caption{SVOM notices broadcast delays for the alert of Sept. 2025}
   \label{fig:notices_broadcast_delay_by_stream}
\end{figure}

Figure~\ref{fig:notices_broadcast_delay_by_instrument} shows the notices broadcast delays for the SVOM alerts of September 2025. The delays are computed as the time difference between the alert message generation on board and the broadcast of the first automated SVOM notice to the SVOM follow-up telescopes, which covers all the processing steps presented in the previous section. We can observe that for more than 50\% of the alerts, the generated notice is broadcasted \textbf{less than 30 seconds} after the alert is detected on board, and all the alerts were broadcasted in less than 15 minutes. The secondary peak centered around 2 minutes is due to alerts that were received in a delayed manner at FSC because they were not transmitted through the VHF network but through the secondary BeiDou communication subsystem which is a bit slower. This is visible in Figure~\ref{fig:notices_broadcast_delay_by_stream}.

\begin{figure}[h]
   \centering
   \includegraphics[width=\linewidth]{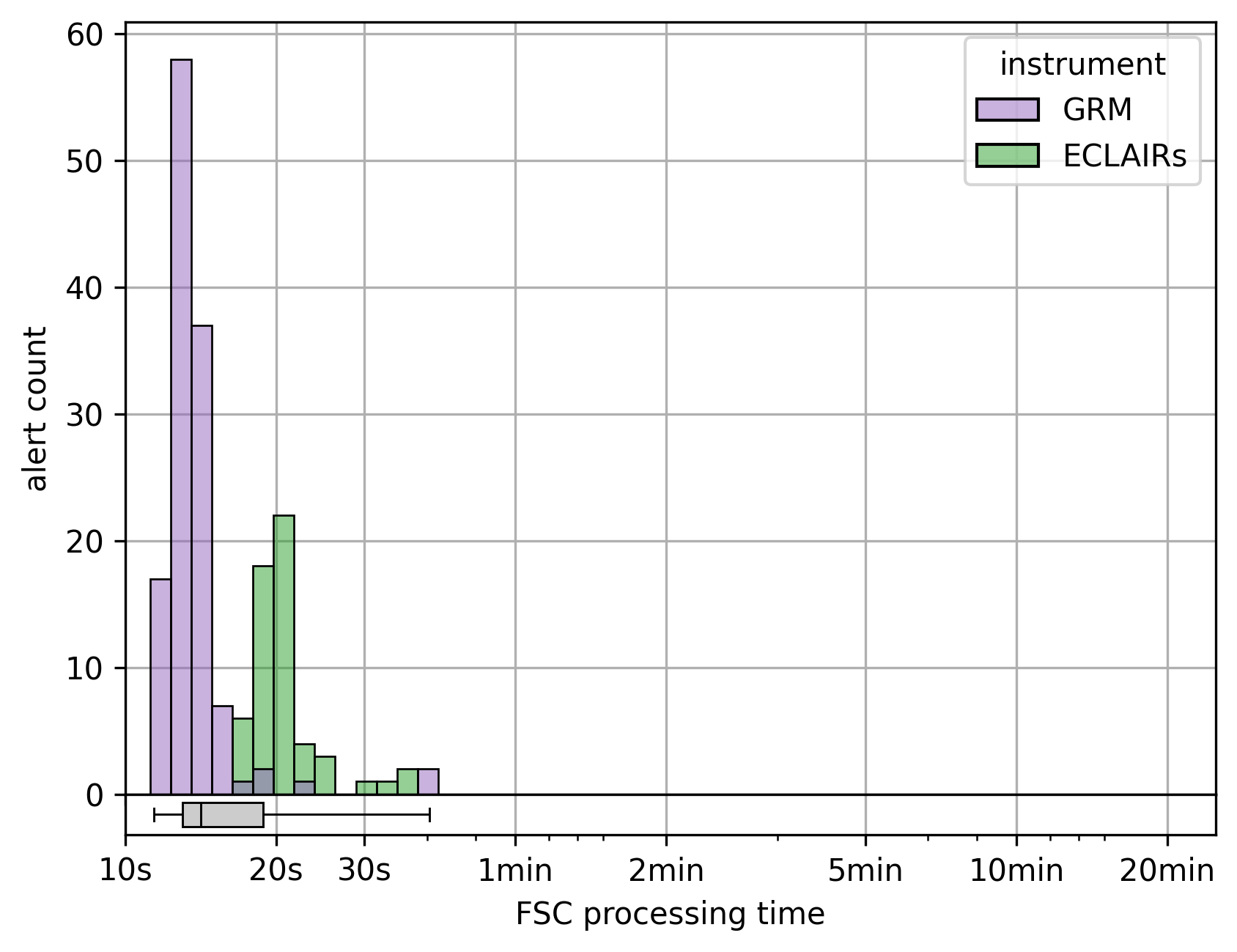}
   \caption{FSC notices processing times for the alert of Sept. 2025}
   \label{fig:notices_fsc_processing_time_by_instrument}
\end{figure}

In order to evaluate the actual FSC processing time and disregard the impact of board-to-ground communication delays, we show on Figure~\ref{fig:notices_fsc_processing_time_by_instrument} the time difference between the FSC reception time of the alert message and the notice broadcast for the alerts displayed on Figure~\ref{fig:notices_broadcast_delay_by_instrument}. The processing time at FSC is very stable and never exceeded 45 seconds, with a median value around 14 seconds. We observe a significant difference between GRM alerts and ECLAIRs alerts which are due to differences in the data processing. The ECLAIRs alerts undergo a few more processing steps before the notice is created and broadcasted, which allow for extra on-ground filtering and automated identification of false alerts.

%
%
\section{Data visualization tools}
\label{sect:dataVisualization}
The current state of FSC and its performances are a reflection of the work of the software development teams participating in the FSC development. Aside from the services already described in the previous sections, a significant amount of software developed for FSC are web-based user interfaces that allow FSC developers, FSC administrators and SVOM scientists to visualize SVOM data, FSC processings and monitor the FSC infrastructure. A total of 23 dedicated user interfaces are developed and maintained at FSC. A few selected example of such interfaces are described in the following sections.

\subsection{FSC website}
\begin{figure}[h]
   \centering
   \includegraphics[width=\linewidth]{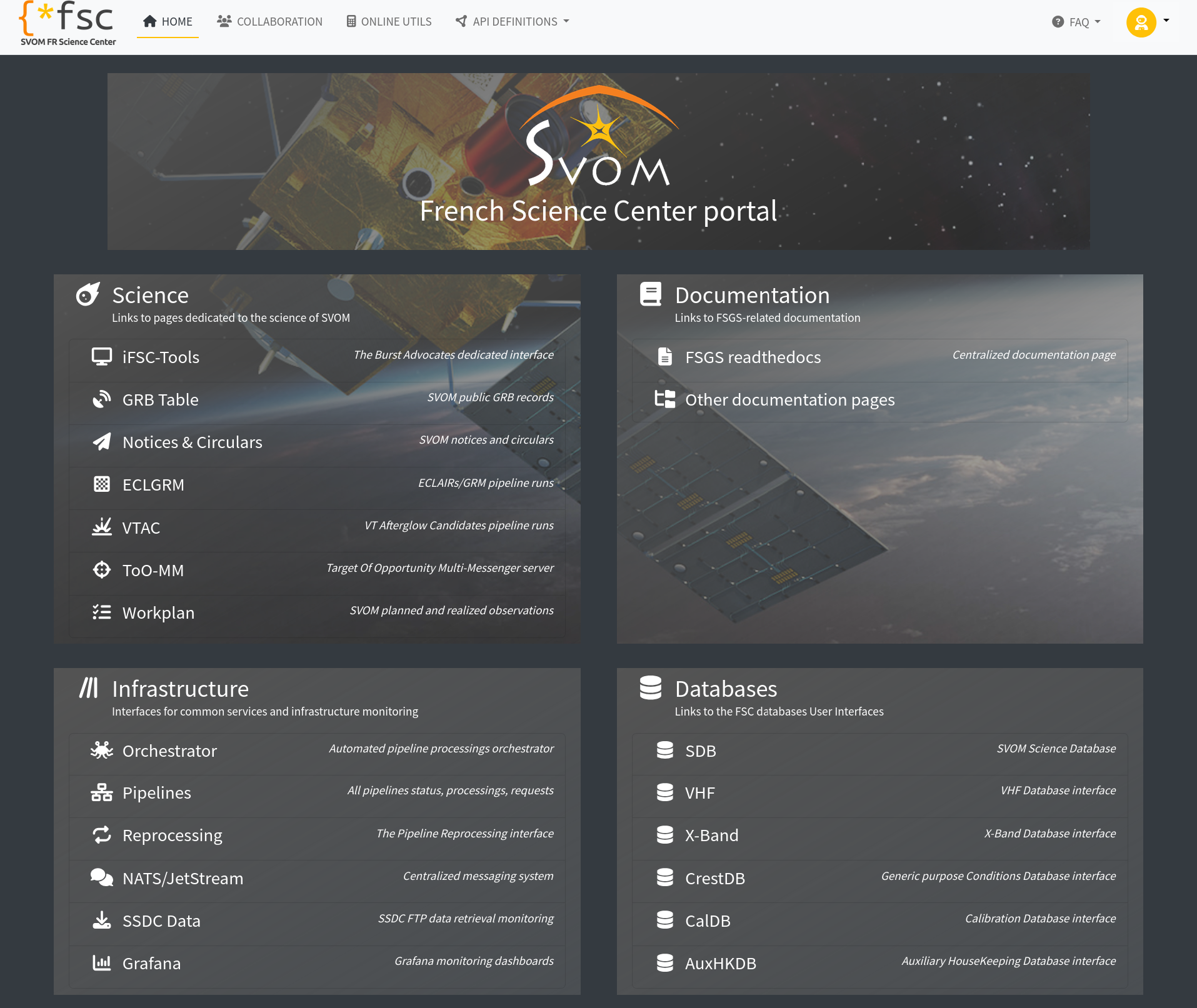}
   \caption{The FSC website home page [internal]}
   \label{fig:fsc_home}
\end{figure}

All the web interfaces developed for FSC operation are gathered under a single proxy with a centralized authentication system provided by a Keycloak\footnote{\url{https://www.keycloak.org/}} instance. All web interfaces are deployed like any other FSC services on the production cloud and are accessible to all SVOM/FSC members through a common portal shown in Figure~\ref{fig:fsc_home}.

\begin{figure}[h]
   \centering
   \includegraphics[width=\linewidth]{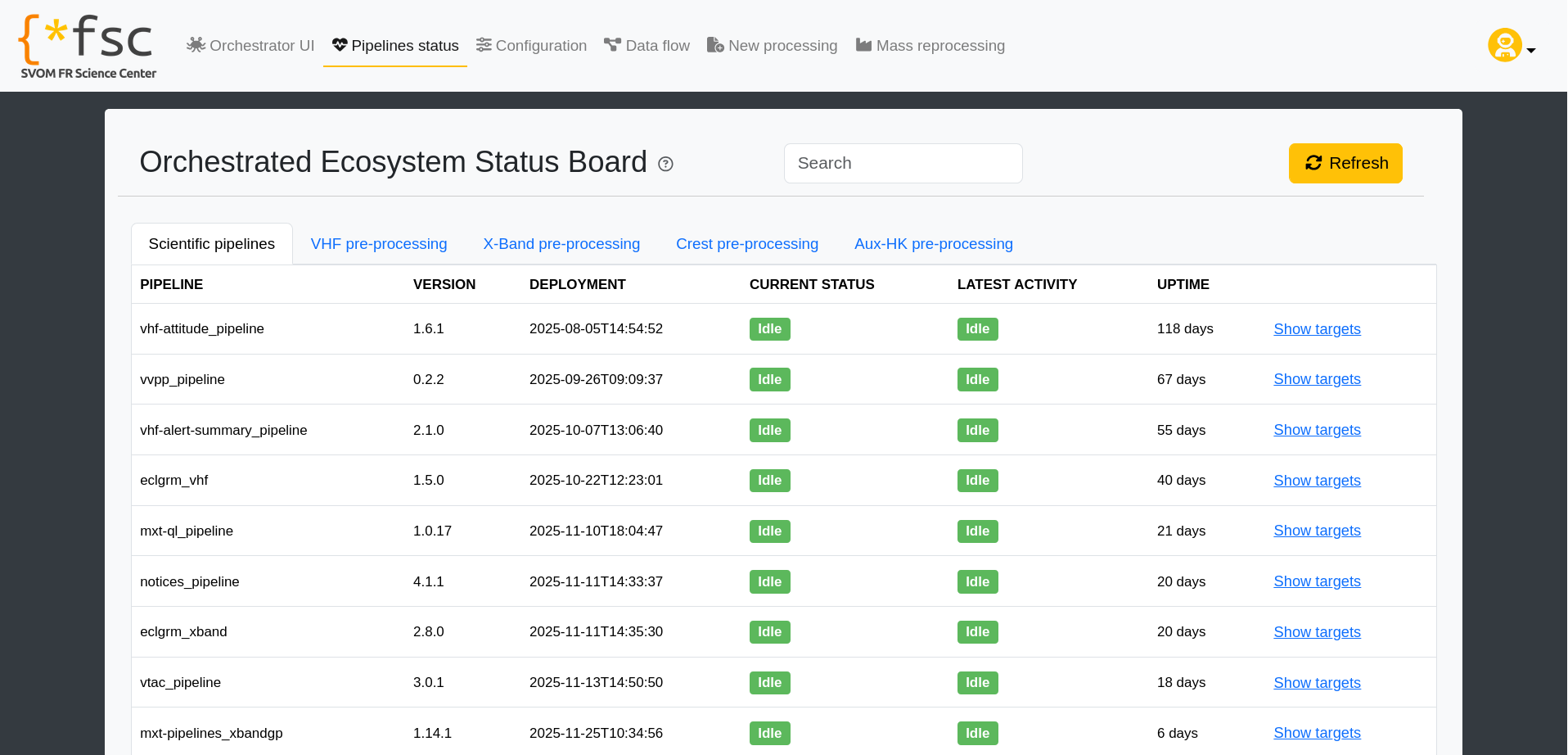}
   \caption{The FSC orchestrator web interface [internal]}
   \label{fig:ui_orchestrator}
\end{figure}

\begin{figure}[h]
   \centering
   \includegraphics[width=\linewidth]{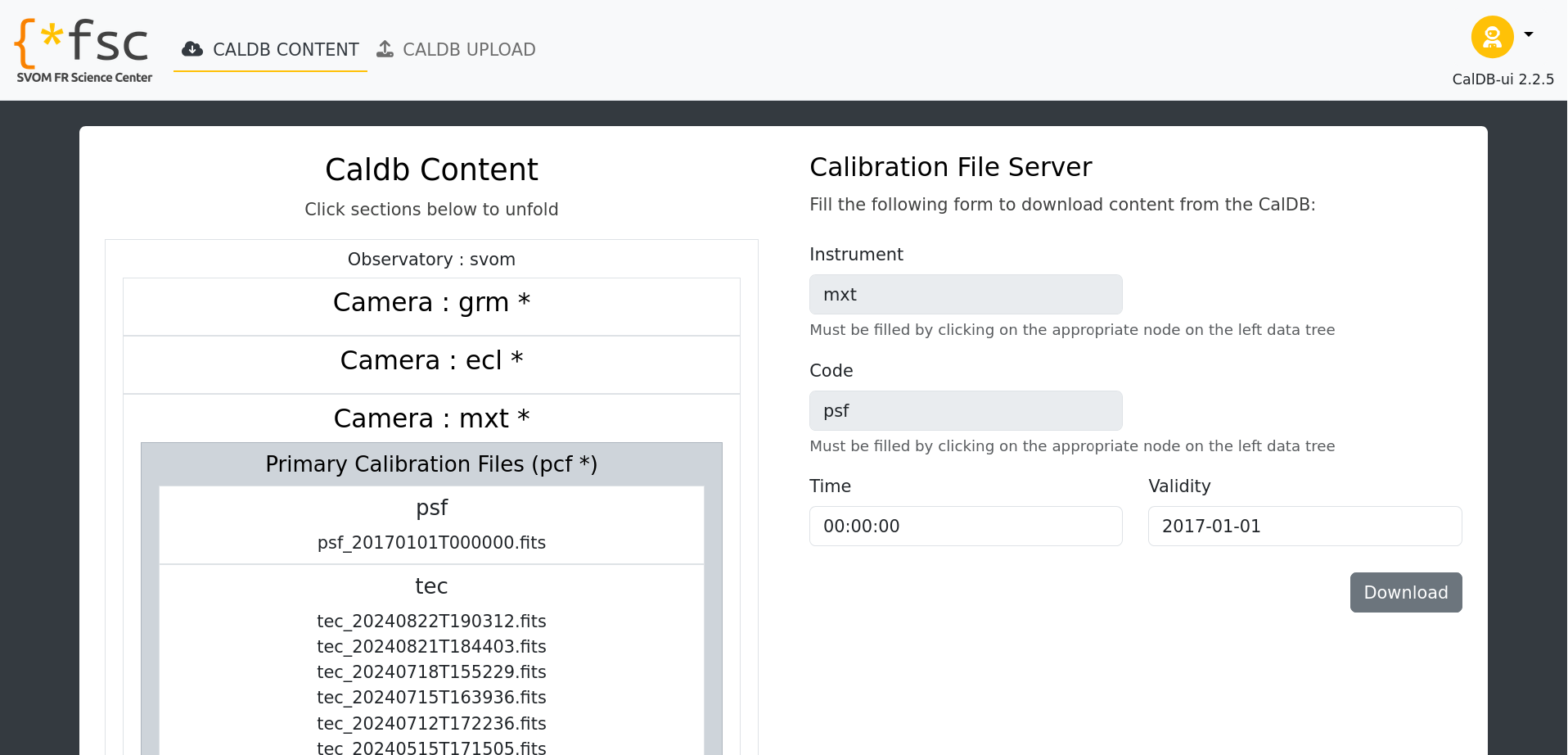}
   \caption{The FSC calibration database web interface [internal]}
   \label{fig:ui_caldb}
\end{figure}

Some of those web interfaces are dedicated to visualize data related to FSC operations and the behavior of common services such as the centralized messaging system or the orchestrator services (see Figure~\ref{fig:ui_orchestrator}). Each FSC database has an associated web interface service that allows to visualize the database content, download data and in some cases and with the proper access rights even upload data (see Figure~\ref{fig:ui_caldb}). Some specialized web pages are focused on SVOM instrument health monitoring by the instrument centers.

\subsection{Infrastructure monitoring interfaces}
\begin{figure}[h]
   \centering
   \includegraphics[width=\linewidth]{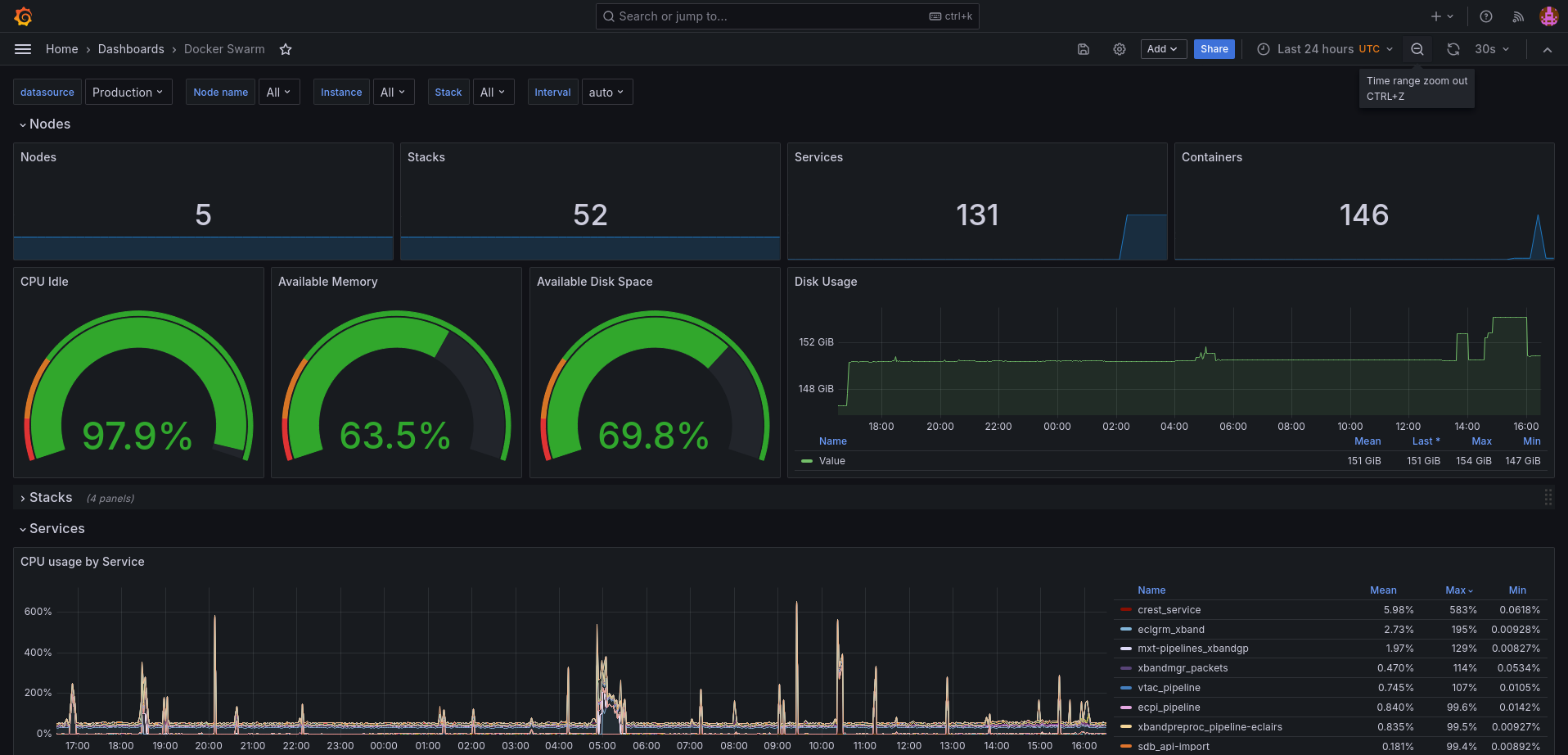}
   \caption{FSC web interface displaying application logs, powered by Grafana [internal]}
   \label{fig:grafana_swarm}
\end{figure}

Most of the FSC user interfaces that are meant to be used by FSC operators and SVOM scientists were specifically developed from scratch to fit the SVOM mission needs. However, the user interfaces used by the FSC administration team to monitor the infrastructure mostly rely on existing tools, such as Grafana\footnote{\url{https://grafana.com/}} which is used for graphical displays of data coming from application logs aggregated using Grafana Loki\footnote{\url{https://grafana.com/docs/loki/latest/}} or time-series system metrics collected using Prometheus\footnote{\url{https://prometheus.io/}}. Figure~\ref{fig:grafana_swarm} shows an example of an FSC Grafana dashboard allowing administrators to monitor the health of the FSC instance and services.

\begin{figure}[h]
   \centering
   \includegraphics[width=\linewidth]{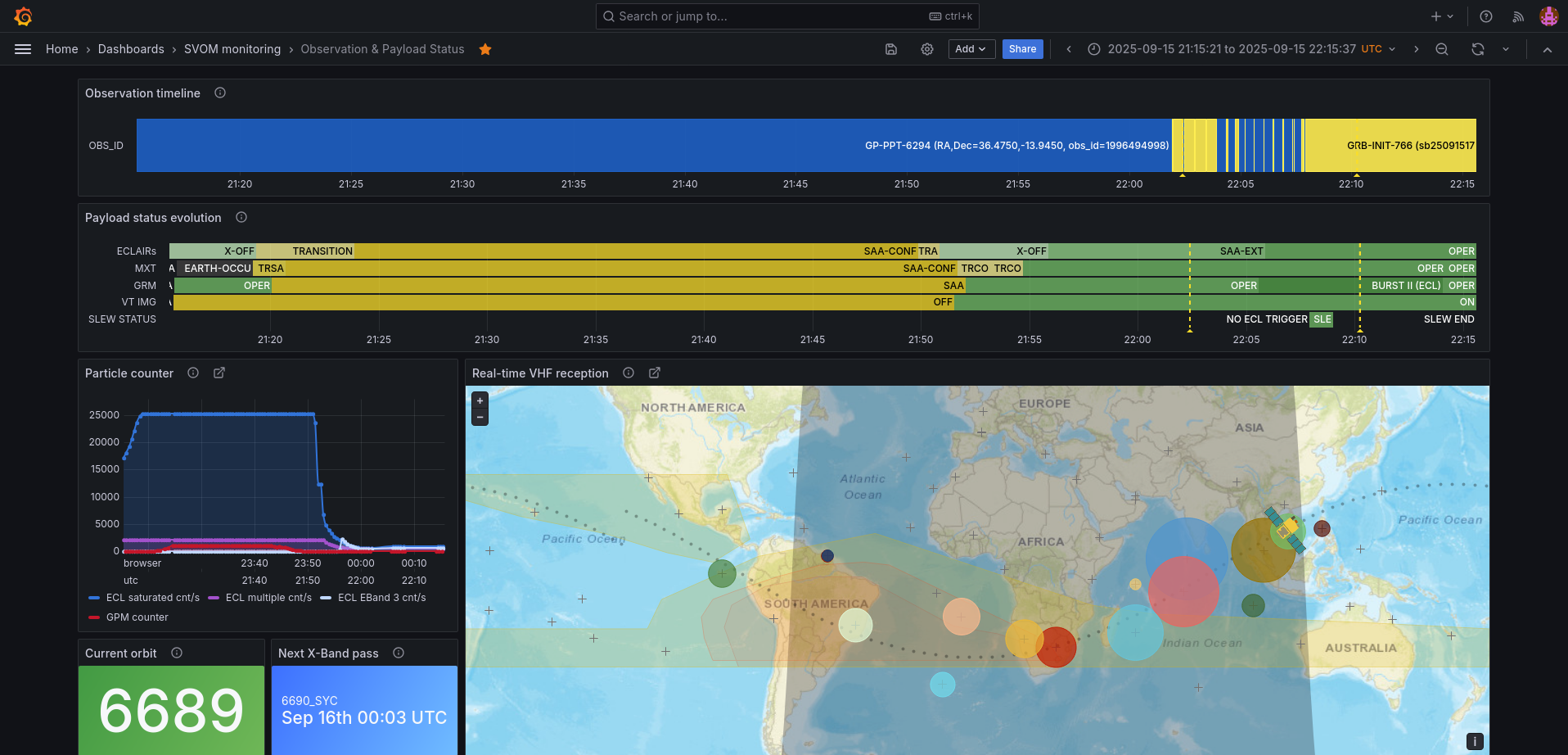}
   \caption{FSC web interface displaying real-time mission monitoring data powered by Grafana [internal]}
   \label{fig:grafana_mission}
\end{figure}

The FSC Grafana instance is also used to display data stored in the various FSC databases, which allows to create real-time monitoring user interfaces for the mission such as illustrated by Figure~\ref{fig:grafana_mission}.

One of the upside of the use of Grafana as a monitoring tool is the use of Grafana "Alert rules" which are used to regularly check specific metrics related to FSC dashboards and fire notifications when threshold are reached. These are used as FSC main alarm system in the event of disk saturation, memory or CPU overload, missing data, services interconnection loss, etc.

\subsection{Science data visualization}
\begin{figure}[h]
   \centering
   \includegraphics[width=\linewidth]{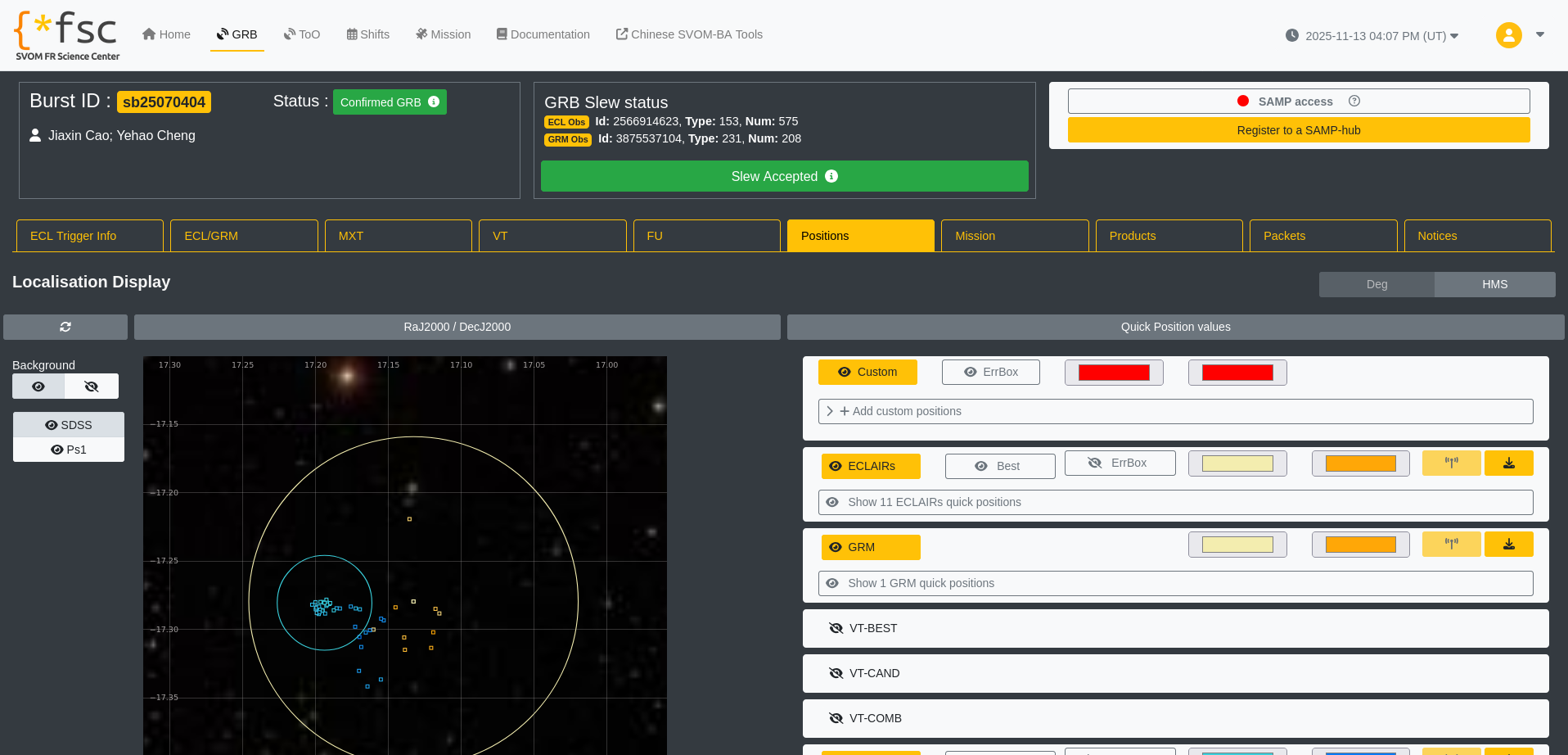}
   \caption{The FSC real-time GRB data visualization web interface, dubbed "iFSC-Tools" [internal]}
   \label{fig:ui_ifsctools}
\end{figure}

The main objective of FSC being to generate scientific products and provide tools and services to access and visualize them, a great deal of web user interfaces developed and maintained at FSC are related to the display of scientific data from the SVOM mission. Those web interfaces and associated tools are described extensively in a dedicated paper \citep{claret+2026+frbatools}. An example of one the central web interfaces for those tasks is shown in Figure~\ref{fig:ui_ifsctools}.

Even if most of the web interfaces developed and maintained at FSC are internal and restricted to the SVOM collaborators, a few of them are also publicly accessible to the scientific community at large. Links to those public pages are available on the FSC homepage at \url{https://fsc.svom.org}.

%
%
\section{Conclusion}
\label{sect:conclusion}
The SVOM French Science Center cloud-based platform is the result of the collaborative work of 10 national French research laboratories, and the central node of the SVOM French Mission Ground Segment. It allows, since the launch of SVOM in June 2024, to perform the archiving, monitoring and scientific analysis of SVOM data for the French community. All the output products and associated visualization tools that were developed for FSC are accessible to all co-investigators of the SVOM collaboration.

The innovative technological choices that were made in developing the FSC proved most efficient in dealing with the flow of SVOM data. With a routine average of 31 thousand automated processings generating 74 thousand scientific products every month, and web interfaces accessible to all SVOM collaborators to access the processed data in real-time, the FSC fulfils its role and allows the scientific needs of the SVOM mission to be met. Thanks to the dedication of the developer teams involved in the FSC development, maintenance and evolution, the FSC already showcases a level of automation that is unprecedented at the beginning of an astrophysical space observatory lifespan.

\section*{Contributors}
CEA/Irfu\inst{1}: A. Claret, M. Bocquier, B. Cordier, D. Corre, N. Dagoneau, A. Formica, D. Götz, L. Jouvin, J-P. LeFèvre, H. Louvin, J. Palmerio, T. Roland, A. Sauvageon, S. Schanne, K. Tazhenova, D. Turpin \\
AIM\inst{2}: T. Sadibekova, J. Rodriguez \\
LAM\inst{3}: F. Agneray, S. Basa, C. Moreau, T. Fenouillet, Y. Roehlly, J-C. Thome \\
ObAS\inst{4}: L. Kleiver , A. Lorang, P. Maggi, L. Michel \\
LUPM\inst{5}: U. Jacob, P. Maeght, F. Piron, T. Maiolino \\
APC\inst{6}:  P. Bacon, N. Bellemont, F. Cangemi, C. Cavet, A. Coleiro, A. Goldwurm, C. Lachaud, S. Le Stum \\
IAP\inst{7}: F. Daigne, L. Domisse, F. Lacreu, G. Tcherniatinsky, J. Wang \\
LUX\inst{8}: Y. Canton, S. Verdani \\
UNIDIA\inst{9}: I. Jegouzo \\
IJCLab\inst{10}: S. Du, N. Leroy, S. Lion, R. Le Montagner, C. Van Hove \\
IRAP\inst{11}: J. Alaux, M. Boiziot, L. Bouchet, O. Godet \\
CPPM\inst{12}: M. Ageron, T. Auphan, H. Benamar, D. Dornic

\begin{acknowledgements}
    The Space-based multi-band astronomical Variable Objects Monitor (SVOM) is a joint Chinese-French mission led by the Chinese National Space Administration (CNSA), the French Space Agency (CNES), and the Chinese Academy of Sciences (CAS).

    The software development activities in all FSC contributing laboratories is supported by CNES.
\end{acknowledgements}

\appendix                  

\bibliography{ms2025-0594.bibtex}

\label{lastpage}

\end{document}